# Performance Analysis Cluster and GPU Computing Environment on Molecular Dynamic Simulation of BRV-1 and REM2 with GROMACS

Heru Suhartanto[1], Arry Yanuar[2] and Ari Wibisono[3]

[1] Faculty of Computer Science, Universitas Indonesia
Depok, 16424, Indonesia

[2] Department of Pharmacy, Faculty of Mathematics and Natural Science,
Universitas IndonesiaDepok, 16424, Indonesia

[3] Faculty of Computer Science, Universitas Indonesia
Depok, 16424, Indonesia

**Abstract**
One of application that needs high performance computing resources is molecular d ynamic. There is some software available that perform molecular dynamic, one of these is a well known GROMACS. Our previous experiment simulating molecular dynamics of Indonesian grown herbal compounds show sufficient speed up on 32 n odes Cluster computing environment. In order to obtain a reliable simulation, one usually needs to run the experiment on the scale of hundred nodes. But this is expensive to develop and maintain. Since the invention of Graphical Processing Units that is also useful for general programming, many applications have been developed to run on this. This paper reports our experiments that evaluate the performance of GROMACS that runs on two different environment, Cluster computing resources and GPU based PCs.

We run the experiment on BRV-1 and REM2 compounds. Four different GPUs are installed on the same type of PCs of quad cores; they are Geforce GTS 250, GTX 465, GTX 470 and Quadro 4000. We build a cluster of 16 nodes based on these four quad cores PCs. The preliminary experiment shows that those run on GTX 470 is the best among the other type of GPUs and as well as the cluster computing resource. A speed up around 11 and 12 is gained, while the cost of computer with GPU is only about 25 percent that of Cluster we built.

## 1. Introduction

Virus is as one of the cause of illness. It is the smallest natural organism. Because of its simplicity and its small size, biologists choose virus as the first effort to simulate life forms with computer, and then choose one of the smallest, called mpsac tobacco satellite for further investigation. Researchers simulate viruses in a d rop of salt water use software – NAMD (Nanoscale Molecular Dynamics) built by the University of Illinois at Urbana-Champaign [13]

Molecular Dynamic (MD) shows molecule structures, movement and function of molecules. MD performs the computation of atom movement in molecular system using molecular mechanics. The dynamic of a protein molecule is affected by protein structure and is an important element of special function and also general function of the protein. The understanding of relation between the dynamical three dimension structures of a protein is very important to know how a protein works. However, further real experiment of protein dynamic is very difficult to be done. Thus, people develop molecular dynamic simulation as a virtual experimental method which is able to analyze the relation between structure and protein dynamic. The simulation explores conformation energy of a protein molecule itself. Up to today, the development of MD simulation is still in progress. MD simulation in general is used to gain information on the movement and the changes of structure conformation of a protein as well as other biological macromolecules. Through this simulation, thermodynamic and kinetic information of a protein can be explored [2].

GROMACS is one of a computer program which is able to run MD simulation and energy minimization. It simulates Newton movement equation of system with hundreds to million molecules. The use of GROMACS can be implemented on biochemistry molecule such as protein which is dynamic and owns complicated binding. In addition, research on non biological molecules, such as polymers can be done with the program [7]. The property of a protein that is dynamic and changing in time is one of the reasons why MD is necessary to be done [6]. With this simulation, the protein molecular movement and interaction that occurs in molecular level on a certain time can be estimated [8].

Our previous experiment using GROMACS on Cluster computing environment produced analysis on MD simulation results with three proteins of RGK sub family; they are Rad, Gem and RemGTpase. MD simulation against protein of subfamily RGK will help us to analyze the activities difference and the regulation mechanism among protein that is member of such subfamily. [9]. As we will mention in the following, we still face some limitation of computing resources.

The parallel computation technique with MD using GROMACS c an visualize atomic details on the formation of s mall DPPC (dipalmitoyl phosphatidyl choline) with time order up to 90 ns [2].

| #CPU | Days | | | | |
|------|------|------|------|------|------|
| 1 | 0 | 41.7 | 206.3 | 1250 | 3750 |
| 4 | 0 | 10.4 | 52.1 | 512.5 | 937.5 |
| 8 | 0 | 5.2 | 26.0 | 156.3 | 468.8 |
| 16 | 0 | 2.6 | 13.0 | 78.1 | 234.4 |
| 32 | 0 | 1.3 | 6.5 | 39.1 | 117.2 |

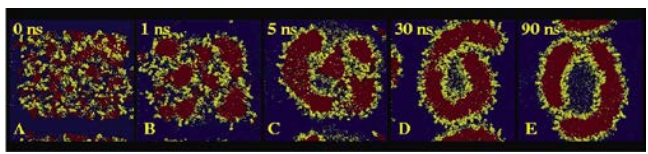

Figure 1. Simulation in 90 ns which shows the formation of DPPC vesicle (dipalmitoyl phosphatidyl choline) [2]

The figure 1 shows that, the simulation requires huge amount of computing resources in the scale of hundreds processors in order to complete the processes within an acceptable short time. However, for some people, to build and maintain a cl uster of hundreds computer acting as hundreds processors requires a lot of other resources such as electricity supplies and space to accommodate the machines. The invention of General Programming o n Graphical Processing Units (GPGPU) which provides hundreds of processors encourage many scientists to investigate the possibility of running their experiments on GPU based computers.

GPU computing with CUDA (Compute Unified Driver Architecture) brings parallel data computation into wider societies. At the end of 2007, there were more than 40 millions application based CUDA GPU developers. In term of price, GPU graphical card is relative cheap. In the year 2007, one unit of the card that is capable of performing 500 G igaFlops costs less than 200 U SD. Currently, one unit of GPU of GEForce GTX 470 that consists of 448 processors cost less than 300 USD. System based on GPU runs relative fast. The bandwidth and the computation reach about 10 times of regular CPU. The micro-benchmark performance reaches mathematical instruction about 472 G FLOPS (for GPU of 8800 U ltra type); the raw bandwidth reaches 86 GB per second (for GPU of Tesla C870 type). Some application can ran faster such as N-body computation that reaches 240 GFLOPS. This is about 12 billion interaction per second. This case study was conducted on molecular dynamic and seismic data processing [2].

The General Purpose Programming on GPU (GPGPU) is the programming processes for general non graphical application on GPU. Initially, the technique is relatively difficult. The problems to be solved should be considered related to graphics. Data has to be mapped into texture maps and the algorithms should be adjusted for image synthesis. Even though this technique is promising, but it is delicate particularly for non graphic developers. There are also some constraints such as the overheads of API graphics, the memory limitation, and the needs for data passes to fulfill the bandwidth.

The libraries for GPGPU is developed and provided in a ready to use programming languages . Many libraries are developed for CUDA such as CUBLAS (Basic Linear Algebra Subprograms in CUDA) and CUFFT (Fast Fourier Transform in CUDA) [3,4]. With the availability of GPGPU, CUDA and the libraries, one of difficulties in developing application run on GPU is solved. With the increasing performance of GPU, it is expected that the GPU computing support the development new application or new techniques in many areas of researches and industries.

Many applications have been developed to run on GPU based computer. The capability of a machine to provide images in highly detailed within a very fast time unit is needed in breast cancer scanning processes. Techniscan, an industry that develops imaging system for automatically ultrasound, switched t heir implementation from CPU based into CUDA and NVIDIA Tesla GPUs. CUDA based system is capable of processing Techniscan algorithm two time faster. Once the appliance obtain operational permit from the government, then patients will know his or her cancer detection results within one visit [5].

On the limited cluster resources that one have and current development of a pplications which run nicely on GPU; motivate us to study the performance of molecular dynamic simulation run on two different computing environment. This paper reports our experiments that evaluate the performance of GROMACS that run on cluster computing resources and GPU based PC.

## 2.The experiment and results

We choose four different types of GPU, namely GeForce GTS 450, GeForce GTX 465, GeForce GTX 470, and Quadro 4000 ( For simplicity, the phrase GeForce will not be mentioned after this). Their specifications are in the following table

Table 1. The specification of PC installed with GPU

| Description | GTX 470 | GTX 465 | Quadro 4000 | GTS 250 |
|---|---|---|---|---|
| Cuda Cores | 448 | 352 | 256 | 128 |
| Memory | 1280 MB GDDR5 | 1024 MB GDDR 5 | 2GB GDDR5 | 1 GB DDR3 |
| Mem. Clock | 1674 MHz | 1603 MHz | | 1100 |
| Mem. interface Width | 320 Bit | 256 Bit | 256 Bit | 256-bit |
| Mem.Bandwidth | 133.9 GB/sec | 102.6 GB/sec | 89,6 GB/sec | 70.4 |

We install GTS 250 into a PC that based on Intel® Pentium ( R ) 4 CPU 3.20 GHz, 4 GB RAM, a 80 GB SATA HDD. While the other GPUs, GTX 465, GTX 470 and Quadro 4000 are installed into PCs with their specification in the following Table 2. We also built 16 cores Cluster computing environment from four PCs with these specification.

Table 2. The specification of PC installed with GPU

| |
|---|
| INTEL Core i5 760 Processor for Desktop Quad Core, 2.8GHz, 8MB Cache, Socket LGA1156 |
| Memory DDR3 2x 2GB, DDR3, PC-10600 Kingston |
| Main board ASUS P7H55M-LX |
| Thermaltake LitePower 700Watt |
| DVD±RW Internal DVD-RW, SATA, Black, 2MB, 22x DVD+R Write SAMSUNG/LITEON |
| Thermaltake VH8000BWS Armor+ MX |
| WESTERN DIGITAL Caviar Black HDD Desktop 3.5 inch SATA 500GB, 7200RPM, SATA II, 64MB Cache, 3.5", include SATA cables and mounting screws |

We installed SDK CUDA 2.3 and some application on our GPU PCs, such as GROMACS 4.0.5 [13] which run on GPU but not fully using processors on the GPU for parallel computation. GROMACS is a program that is designed to run serially or in parallel. The parallel features in GROMACS are provided to speed up the simulation processes. GROMACS has parallel algorithm related to domain decomposition. The domain decomposition algorithm is useful to solve boundary value problems by decomposing the main problem into some smaller boundary value problems. In GROMACS, the domain decomposition algorithm divides the simulation works into different processors so that the processes can be done in parallel. Each of the processor will collect and coordinate the motion of simulated particles.

In the parallel simulation processes, processors will communicate to each others. Workload imbalance is caused by different particle distribution on each of the processors as well as the different particle interaction among the processors. In order to have work balance, GROMACS uses dynamic load balancing algorithm where the volume of each domain decomposition can be adjusted independently . The *mdrun* script of GROMACS will automatically start dynamic load balancing when there is instability energy computation within 5 percent or more. Load imbalance is recorded on output log created by GROMACS.

As there is no test performance record yet for GROMACS 4.5 [13] running on Quadro 4000, so when a user runs MD simulation, a warning message appear saying that the graphic card used has not been tested to use GROMACS 4.5 software. A further investigation is necessary in order to see how GROMACS can run nice much better on Quadro 4000.

In the experiment, we use two different compounds. Breda Virus 1 abbreviated as BRV-1 or 1 br v (***PDB id 1BRV***) [14]. Its scientific name is B ovine respiratory syncytial virus [*Biochemistry, 1996, 35 (47), pp 14684–14688*]. The second one is REM2 (***PDB id: 3CBQ***). REM2 is the molecular function induces angiogenesis (essential for growth, invasion and metastasis tumor). So REM2 a potential target to overcome condition of angiogenesis. REM2 was known to regulate p53 to the nature of immortal in somatic cells. REM2 also cause the stem cells immortal, REM2 alleged role in the mechanism self-renewal in stem cells to hESC (human Embryonic Stem Cell which protects from the process of apoptosis (programmed cell death). [12]

In the following table 3, we provide the time required in the simulation of 1BRV compound in various values of time steps

Table 3. The time of simulation processes of 1 BRV

| | Time steps | | | | |
|---|---|---|---|---|---|
| Resources | 200ps | 400ps | 600ps | 800ps | 1000ps |
| Cluster 16 | 39m:22s | 1h:19m:13s | 1h:59m:32s | 2h:39m:37s | 3h:19m:08s |
| GTS 250 | 23m:26s | 46m:57s | 1h:10m:23s | 1h:33m:58s | 1h:57m:18s |
| Quaddro 4000 | 6m:45s | 13m:35s | 20:23 | 27m:05s | 33m:29s |
| GTX 465 | 8m:48s | 8m:32s | 12m:46s | 16m:57s | 21m:21s |
| GTX 470 | 3m:30s | 7m:00 | 10m:29s | 14m:07s | 17m:14s |

It is obvious that the experiments with GTX 470 outperform simulations in other environment. For example, in 1000 ps time steps with GTX 470, it requires 17 minutes and 14 second to finish the simulation, while with GTS 250, it requires 1 hour 27 minutes and 32 second which is almost seven times of GTX 470. The performance degrades from GTX 470, GTX 465, Quadro 4000, and to GTS 250. This perhaps that each of these GPUs has different number of GPU cores, 448, 352, 256 and 128 cores respectively. The interesting one is that all the experiment on GPU outperforms those of Cluster 16.

We next simulate REM2, and the time spent in simulation is provided in the following table 4. It is also obvious that the similar pattern happen in this simulation that GTX 470 performs best, and all GPU outperform those of Cluster 16.

Table 4 The time of simulation processes of REM2

| Resources | Time steps | | | | |
|---|---|---|---|---|---|
| | 200ps | 400ps | 600ps | 800ps | 1000ps |
| Cluster 16 | 45m:21s | 1h:27m:27s | 2h:13m:05s | 2h:57m:01s | 3h:43m:15s |
| GTS 250 | 27m:56s | 56m:11s | 1h:24m:26s | 1h:52m:23s | 2h:23m:32s |
| Quaddro 4000 | 6m:45s | 13m:35s | 20m:23s | 27m:05s | 33m:29s |
| GTX 465 | 4m:13s | 8m:32s | 12m:46s | 16m:57s | 21m:21s |
| GTX 470 | 3m:30s | 7m:00s | 10m:29s | 14m:07s | 17m:14s |

We finally provide the speed up of GTX 470 compare with Cluster 16, the following table 5 summarizes the result from each simulation of 1BRV (SU-1BRV) and REM2 (SU-REM2).

Table 5. Speed up of GTX 470 relative to Cluster 16

| SU\tsteps | 200ps | 400ps | 600ps | 800ps | 1000ps |
|---|---|---|---|---|---|
| SU-REM2 | 12.95714 | 12.49286 | 12.69475 | 12.53955 | 12.95455 |
| SU-1BRV | 11.24762 | 11.31667 | 11.40223 | 11.30697 | 11.55513 |

It is obvious that GTX470 gain speed up at about 11 to Cluster16 on 1BRV simulation, and about 12 on REM2 simulation. In term of hardware price, the PC with GTX 470 is about 25 percent that of Cluster 16 with the same specification.

## 3. Conclusion

Molecular dynamics simulation using BRV-1 and REM2 is conducted on two different computing environment, cluster computing of 16 nodes and GPU computing of various types of NVIDIA – GTS 250, GTX 465, GTX 470 and Quaddro 4000. Both cluster and GPU computing has the same hardware specification, except that of GTS 250. The experiment show the efficacy of GTX 470 compare to the others. It gain speed up around 11 and 12, the cost of this GTX 470 computer is only about 25 percent that of Cluster 16 processors. Even though our preliminary findings are interesting, but it needs further investigation as the development of Molecular Dynamics codes on GPU is still in progress.

## Acknowledgments

This research is supported by Universitas Indonesia Multi Discipline Research Grant in the year 2010. The Final stage of the writing process of the paper was conducted when the first author visited the School of ITEE – the University of Queensland.

**Heru Suhartanto** is a Professor in Faculty of Computer Science, Universitas Indonesia (Fasilkom UI). He has been with Fasilkom UI since 1986. Previously he held some positions such as Post doctoral fellow at Advanced Computational Modelling Centre, the University of Queensland, Australia in 1998 – 2000; two periods vice Dean for General Affair at Fasilkom UI since 2000. He graduated from undergraduate study at Department of Mathematics, UI in 1986. He holds Master of Science, from Department of Computer Science, The University of Toronto, Canada since 1990. He also holds Ph.D in Parallel Computing from Department of Mathematics, The University of Queensland since 1998. His main research interests are Numerical, Parallel, Cloud and Grid computing. He is also a member of reviewer of several referred international journal such as journal of Computational and Applied Mathematics, International Journal of Computer Mathematics, and Journal of Universal Computer Science. Furthermore, he has supervised some Master and PhD students; he has won some research grants; holds several software copyrights; published a number of books in Indonesian and international papers in proceeding and journal. He is also member of IEEE and ACM.

**Arry Yanuar** is an assistant Professor of Department Pharmacy, Universitas Indonesia since 1990. He graduated from undergraduate program Department of Pharmacy, Universitas Indonesia in 1990. He also holds Apoteker Profession certificate in 1991. In 1997, he finished his Master Program from Department of Pharmacy, Universitas Gadjah Mada. He holds PhD in 2006 from Nara Institute of Science and Technology (NAIST), Jepang, with Structure Biology/protein Cristalography laboratorium. In 1999-2003 he worked as pharmacy expert ini ISO certification for pharmacy industries at Llyod Register Quality Assurance. In 2002, he visited National Institute of Health (NIH), Bethesda, USA. He won several research grants and published some paper at international journals and conferences.

**Ari Wibisono** is a research Assistant at Faculty of Computer Science Universitas Indonesia (Fasilkom UI). He graduated from undergraduate program Fasilkom UI and currently takes Master program at Fasilkom UI.